\documentclass[twocolumn,showpacs,preprintnumbers,amsmath,amssymb]{revtex4-1}
\usepackage{epsfig}
\usepackage{graphicx}
\usepackage{epsfig}
\usepackage{color}
\usepackage{rotating}
\usepackage{amssymb}
\usepackage{amsmath}
\usepackage{graphicx}

\def\beq{\begin{eqnarray}}
\def\eeq{\end{eqnarray}}

\newcommand{\be}{\begin{equation}}
\newcommand{\ee}{\end{equation}}
\newcommand{\bea}{\begin{eqnarray}}
\newcommand{\eea}{\end{eqnarray}}

\begin{document}

\title{The stochastic thermodynamics of a rotating Brownian particle in a gradient flow}
\author{Yueheng Lan}
\email{lanyh@mail.tsinghua.edu.cn}
\affiliation{The Department of Physics\\
Tsinghua University, 100084 Beijing, China}
\author{Erik Aurell}
 \altaffiliation[Also at ]{Dept. Information and Computer Science and Aalto Science Institute (AScI),
Aalto University, Espoo, Finland
}
 \email{eaurell@kth.se}
\affiliation{Dept.~Computational Biology and ACCESS Linnaeus Centre, KTH-Royal Institute of Technology,
106 91 Stockholm, Sweden}

\begin{abstract}
We compute the entropy production engendered in the environment from
a single Brownian particle which moves in a mean flow, and show that
it corresponds in expectation to classical near-equilibrium
entropy production in the surrounding fluid with specific mesoscopic transport coefficients.  With temperature 
gradient, extra terms are found which results from the nonlinear interaction between the particle and
the non-equilibrated environment. 
The calculations are carried out in the multi-time formalism and in the advection-dissipation limit
where the Stokes number $\hbox{St}$ of the flow tends to zero and the P\'eclet number $\hbox{Pe}$
diverges but the combination $\hbox{St}\cdot\hbox{Pe}$ remains constant.
\end{abstract}

\pacs{05.10.Gg,05.40.Jc,05.60.Cd,05.70.Ln}
\keywords{Stochastic thermodynamics, Brownian particle, entropy production, advection-diffusion limit}

\maketitle


{\it Introduction:}
Motion of micro-sized particles in a non-equilibrium environment has recently inspired
interest among researchers in diverse fields of science and
engineering~\cite{Esposito09-review,dhont-colloid,hub72br,ch13br,han06han,fak10br}. Interesting
observables, such as diffusion constant, correlation length and various transport coefficients,
are measured or computed to understand such processes. In all this characterization, entropy
production is a key to the study of energy dissipation, which however could not be fully
described by the commonly used over-damped Langevin description~\cite{Celani12}. The interaction
of the inertia of the particle with the non-equilibrated surroundings has to be carefully
taken into account.

The change of entropy per unit time in a macroscopic
fluid due to dissipative processes is a quantity defined when the fluid is close to local
thermal equilibrium and which depends on gradients of intensive quantities such as
temperature and local (mean) velocity~\cite{GrootMazur-book,LandauLifshitz-vol6}.
On the other hand, entropy production in the environment has emerged as a fundamental quantity
in mesoscopic physics underlying fluctuation relations which hold both close to and far from
equilibrium~\cite{Searles08-review,Jarzynski11-review,seifert12-review}.
This entropy production is mathematically the logarithm of the ratio of
the probabilities to observe a forward and reversed system trajectory, and is therefore
a functional of the whole system history~\cite{Kurchan98,LS99,ChG07}.
The modern notion of far-from-equilbrium entropy
production should be put into relation with the traditional near-to-equilibrium concept,
and a first step in this direction was taken in~\cite{Celani12},
where it was shown that for an overdamped Brownian particle in a changing environment
there is an ``anomalous'' contribution to the entropy production (in the second sense)
which reads
\begin{equation}
\delta S_{\hbox{\small anom}}[x(t)_{t_i}^{t_f}] = \int_{t_i}^{t_f} \frac{5T}{6\gamma} \frac{(\nabla T)^2}{T^2} d
\label{eq:S-anom}
\end{equation}
where $x(t)$ is the particle trajectory, $T$ is the temperature and $\gamma$ is the friction coefficient.
Anomalous means that this contribution cannot be referred to a functional of the overdamped
motion in space, and therefore represents an entropy production which
belongs to the surrounding medium.
The expected entropy production per unit time from an ensemble of non-interacting
Brownian particles with time-independent number density $\rho$ is then
\begin{equation}
\frac{d}{dt_f}<\delta S_{\hbox{\small anom}}[x(t)_{t_i}^{t_f}]> = \int \frac{5T\rho}{6\gamma} \frac{(\nabla T)^2}{T^2} dV
\label{eq:S-anom-average}
\end{equation}
This agrees with the normal form of the
entropy production (in the first sense) in a fluid at rest in a
temperature gradient, as given in eq. 49.6 of~\cite{LandauLifshitz-vol6},
with a mesoscopic thermal conductivity $\kappa^{(th)}=\frac{5T\rho}{6\gamma}$.
In this contribution we extend these results in two directions.
First we show that if fluid has a spatially varying mean velocity then there
is an anomalous entropy production which for a point particle
and in a constant temperature field reads
\begin{equation}
\delta S_{\hbox{\small anom}}^{(u)}[x(t)_{t_i}^{t_f}] =\int_{t_i}^{t_f} \frac{m}{2\gamma}(\partial_ju^i\partial_iu^j+\partial_ju^i\partial_ju^i) dt
\label{eq:S-anom-u}
\,,
\end{equation}
where $m$ is the mass of the diffusing particle.
As the last two terms in eq. 49.6 of~\cite{LandauLifshitz-vol6} equation~(\ref{eq:S-anom-u}) represents
hydrodynamic dissipation. 
%
Secondly we show that a Brownian particle of finite extent and having rotational
degrees of freedom generates additional anomalous entropy production taking the
same functional form as (\ref{eq:S-anom}). If the particle is spherical
(angular rotation frictional matrix and moment of inertia tensor both proportional
to identity) then the new contribution to the thermal conductivity is $\kappa^{(th)}_{\hbox{rot}}=\frac{3T\rho}{2(2\gamma_2+\gamma)}$
where $\gamma_2$ is an angular friction coefficient.

The technique used to establish these results is asymptotic expansions
using the multi-time formalism~\cite{BenderOrszag-book,BLP78,PavliotisStuart08}.
To arrive at a non-trivial dependence on mean flow, \textit{i.e.}
the result (\ref{eq:S-anom-u}), we need to take
an advection-diffusion limit which mixes a conservation law on a faster time
scale with proper dissipative action on a longer time scale, for
earlier uses of analogous techniques in other contexts, \textit{see}~\cite{Frisch87}
and \cite{MartinsMazzinoPMG12}.
The more general case, a rotating non-symmetric object coupled to a mean flow in
particular, are computationally somewhat involved, and hence presented in Supplementary Information.

{\it Dimensions, time scales and basic equations:}
In the framework of Stochastic Thermodynamics it is assumed
that the time scales of the surrounding fluid are much faster than
those of the object. The translational degrees of freedom obey the Kramers-Langevin equations
\begin{equation}
m\frac{d(u^i+v^i)}{dt} = -\gamma v^i + f^i +\sqrt{2T\gamma}\dot{\eta^i}\quad \frac{dx^i}{dt}=u^i+v^i
\label{eq:Langevin Kramers}
\end{equation}
where $f^i$ is an external force (protocol), $u$ is the mean flow,
$T$ is the temperature in units such that $k_B=1$, and $\gamma$ is a friction coeffient.
Without mean flow the rotations of the body are described by Euler equations supplemented by
angular friction and angular noise~\cite{dhont-colloid,hub72br}
\begin{equation}
\frac{dQ^{\alpha}_{\beta}I^{\beta}_{\sigma}\omega^{\sigma}}{dt} =
-Q^{\alpha}_{\beta}\Gamma^{\beta}_{\sigma} \omega^{\sigma}+M^{\alpha}+\sqrt{2T}Q^{\alpha}_{\beta}\Sigma^{\beta}_{\sigma}
\dot{\xi}^{\sigma} \,. \label{eq:wi-simple}
\end{equation}
where $\omega$ is the angular velocity in a coordinate system fixed in the body, $M$ is a an external torque
in the laboratory coordinate system, $I$ and $\Gamma$ are the moment of inertia tensor and rotational friction matrix, $\Sigma$ the strength of
angular momentum noise, and $\Gamma^{\alpha\beta}=\sum_{\sigma}\Sigma^{\alpha}_{\sigma}\Sigma^{\beta}_{\sigma}$.
$Q$ is the rotation matrix from the body frame of reference to the laboratory.
The noise sources are assumed delta-correlated.
As we will see it is a consistent approximation to ignore the effects of mean flow
on rotation. The effects of combining rotation with
a non-isotropic mobility tensor ($\gamma$ in (\ref{eq:Langevin Kramers})
promoted to a matrix) will be reported elsewhere~\cite{Marino-in-preparation}.

To give numbers, assume the Brownian particle to be something
like a polystyrene ball of radius $R \sim 1 \mu m$ moving in water near
room temperature. Several different time scales exist across the
microscopic and the mesoscopic regimes. The shortest is the microscopic collision time
which is $t_c \sim 10^{-12} s$ while the shortest time we will be interested in is the momentum
relaxation time, which we can estimate from Stoke's law to be about $t_r \sim 10^{-7}s$.
The relaxation time and the thermal velocity of the object $v_{\mbox{th}}=\sqrt{2T/m}\sim 1 mm/s$
combine to give a small spatial scale $\ell=t_r v_{th}\sim 10^{-10}m$. We assume that temperature, friction coefficients,
external force and external torque as well as (see below) mean flow vary on a spatial scale $L$ which
is much larger than $\ell$. Their ratio $\ell/L=\epsilon$ is then a small parameter. The diffusion coefficient
$D=\frac{T}{\gamma}$ would be on the order of or less than $1 (\mu m)^2/s$ giving a slow (diffusive) time scale $L^2/D$.
The potential of the external force is assumed comparable to thermal energy, which implies $f\sim T/L$.
To include rotations we first note that the relaxation of angular velocity happens on the same time scale $t_r$ since
$I\sim mR^2$ and $\Gamma\sim \gamma R^2$ \cite{spinning-torque,bird02trans}. The typical (thermal) angular velocity
is about $10^3$ radians/s such that the object rotates about $10^{-4}$ radians in a time $t_r$.
Orientation of the object will diffuse one radian on a time scale $\frac{1}{T/\Gamma}\sim R^2/D$ which is of the same
order as it takes for the particle to move the distance of its own radius.

We here and in the following need to formally take $R$ of the same order as $L$
in which case the slow time scale $t_f=L^2/D=R^2/D$ is on the order of seconds.
The more realistic assumption of $R<<L$ would bring in two spatial scales ($R$ and $L$)
and two different long times ($R^2/D$ and $L^2/D$), and would need to be treated by
techniques such as those developed in~\cite{MazzinoMusacchioVulpiani05} and~\cite{MartinsMazzinoPMG12},
outside the scope of the present work.

{\it The advection-diffusion limit of a Brownian particle in a mean flow:}
We consider first the effects of mean flow and focus on a point particle (no rotation).
A constant mean flow can be eliminated by a change of reference, and we can therefore further assume
zero average mean flow over the large spatial scale $L$. The variation of $u$ over distances
$L$ hence defines a characteric mean flow amplitude $\overline{u}$ and a corresponding time $t_u=L/\overline{u}$.
The ratio $\hbox{St}=\frac{t_r}{t_u}=\frac{t_r \overline{u}}{L}$ is the Stokes number
of the particle in the flow, $\hbox{Pe}=\frac{t_f}{t_u}=\frac{L \overline{u}}{D}$ is the P\'eclet number,
and $\hbox{St}\hbox{Pe} = \overline{u}^2/v_{th}^2$.
Our basic scaling assumption is that the ratios $\frac{\ell}{L}=\epsilon$ and $\frac{t_r}{t_f}=\frac{Pe}{St}=\epsilon^2$ are small.
The effects of the mean flow however depend on how $\hbox{St}$ and $\hbox{Pe}$ separately scale with $\epsilon$.

A first possibility is the overdamped limit when $\hbox{St}\to 0$ and $\hbox{Pe}$ constant.
The kinetic energy of the mean flow is then small compared to thermal energy and external potential energy,
and there are basically no effects on the time and length scales we consider.
A second possibility is $\hbox{Pe}\to \infty$ and $\hbox{St}$ constant which is the case of inertial
particles moving in a velocity field $u$~\cite{Maxey87,BalkovskyFalkovichFouxon01}.
In this case diffusion can be considered weak, the anomalous entropy production terms from the fluctuating velocity
would be negligible, but there would instead be entropy production terms
from coarse-graining in space~\cite{Esposito12,Bo14}.
As noted above these are interesting questions, but outside the scope of the present work.

We here consider the third possibility when  $\hbox{St}\sim \hbox{Pe}^{-1}$,
which means that the kinetic energy of the mean flow is comparable to the thermal energy
and $t_u\sim t_r\epsilon^{-1}$.
Continuing on the example above we can imagine a
mean flow to be generated in the annulus between two rotating cylinders
of radii $r_-$ and $r_+$ imparting tangential velocities $u_-$ and $u_+$ to the liquid.
If the two cylinders have radii about 1 cm and the width of the annulus is about 1 mm the
assumption of scale separation (neglecting rotation) is easily satisfied since $\ell/L=\epsilon\sim 10^{-7}<<1$.
Similarly, the condition that the $\overline{u}$ should be of the order of
the thermal velocity (about $1$ mm/s) of the particles means that the angular velocities of the
two annuli only need to differ by about 1 rpm.
The assumption that the mean
flow $u$ has no structure on spatial scales smaller than $L$ on the other hand
places a limit on how large $L$ (and hence the scale separation) can be, as it supposes that a Reynolds
number built on $\overline{u}$ and $L$ is sub-critical~\cite{Re-crit}.
To the stochastic equations (\ref{eq:Langevin Kramers})
corresponds a Fokker-Planck equation
in a probability density $P(x,v,t)$ over positions and velocities.
In the multi-scale expansion we posit two scales in space as above and three scales in time $t_0=t_r$, $t_1=t_r/\epsilon$
and $t_2=t_r/\epsilon^2=t_f$, and assume that $P$ depends separately on all the scales and can be expanded
as
\begin{equation}
P=P^{(0)}+\epsilon P^{(1)} + \epsilon^2 P^{(2)} + \ldots
 \end{equation}
In the advection-diffusion limit $t_1$ is comparable to $t_u$. The left hand side of the
the first equation in (\ref{eq:Langevin Kramers})
can be simply written as $m\frac{dv^i}{dt}$ if in the right hand side
we change the force $f^i$ to $f^i_{eff} -mv^j\partial_j u^i$,
where the ``effective force'' is $f^i - m\partial_t u^i - mu^j\partial_j u_i$.
The hierarchy of equations to be
solved are thus
\begin{equation}
\left(\partial_{t_0}+\epsilon \partial_{t_1} + \epsilon^2 \partial_{t_2}\right)P = \left(M^{\dag}+\epsilon L^{\dag} + \ldots\right) P
\end{equation}
where $M^{\dag}=\frac{\gamma}{m}\left(\partial_{v_i}(v_i P)+\frac{T}{m}\partial^2_{vv}P\right)$
and $\epsilon L^{\dag} + \ldots$ are higher-order terms. To order $\epsilon^k$ we have
$(\partial_{t_0}+M^{\dag})P^{(k)}$ equal to terms dependent on $P^{(0)}\ldots P^{(k-1)}$
which entails the solvability condition that the right hand sides are
orthogonal to functions constant in $v$~\cite{Celani12}.
To order $\epsilon$ the solvability condition yields the conservation law
$\partial_{t_u}\rho_0 + \partial_{x_i}(u^i\rho_0)=0$,
where $\rho_0=\int dv P_0$~\cite{order}.
On order $\epsilon^2$ one gets that
the same terms for a first order spatial density $\rho_1=\int dv P_1$, \textit{i.e.}
$\partial_{t_u}\rho_1 + \partial_{x_i}(u^i\rho_1)$, together with diffusive
terms for $\rho_0$ vanish. These diffusive terms are, disregarding variations of $T$ and $\gamma$
in space,
$\partial_{t_f}\rho_0 + \frac{1}{\gamma} \partial_{X_i}(f^i_{eff}\rho_0) -\frac{T}{\gamma}\partial^2_{X_iX_i}\rho_0$
where the effective force $f^i_{eff}$ was introduced above.
The combinination  $\tilde\rho_0=\rho_0+\epsilon \rho_1$ therefore obeys, up to terms of order $\epsilon^3$,
the same equation as the Fokker-Planck equation of the process
\begin{equation}
dX^i_t = u^idt + \frac{1}{\gamma}f^i_{eff}dt + \sqrt{\frac{2T}{\gamma}}\circ dW_t^i
\label{eq:overdamped-space-meanflow}
\end{equation}
This equation explains the term ``advection-diffusion limit'', where a faster process $dX^{i,(1)}_t = u^idt$ on time scale $t_u$ is
mixed with a slower process $dX^{i,(2)}_t = \frac{1}{\gamma}f^i_{eff}dt + \sqrt{\frac{2T}{\gamma}}\circ dW_t^i$ on time
scale $t_f$. One may note the appearence of the Maxey terms
$-\frac{m}{\gamma}\left(\partial_t u^i + u^j\partial_j u_i\right)$ of inertial particle theory~\cite{Maxey87}.
In the case that $T$ and $\gamma$ depend on space the diffusive term in
(\ref{eq:overdamped-space-meanflow}) should be corrected by adding the ``spurious'' terms
$\frac{1}{\gamma}\left(-\frac{1}{2}\partial_i T -\frac{T}{2}\partial_i\log\gamma\right)dt$,
see ~\cite{MatsuoSasa00} and \cite{Celani12}.

{\it The anomalous entropy production in a mean flow:}
The logarithm of the ratio of
the probabilities to observe a forward and reversed system trajectory
defined by Eq.~(\ref{eq:Langevin Kramers}) is,
as we show in Supplementary Material for the more general case including rotations,
\begin{eqnarray}
S_{env} &=& \int \frac{(f^i_{eff}-mv^j\partial_j u^i)v^i}{T}dt-\frac{mv}{T}\circ dv
\label{eq:S-ent}
\end{eqnarray}
where $f_{eff}$ was defined above. Using the Sekimoto sign convention~\cite{Sekimoto-book}
the heat exchange $\delta Q$ is $-TS_{env}$ which implies that an infinitessimal 
heat $d'Q=mv\circ d(u+v) + v\partial_x Vdt$. A first law on the level of trajectories
is hence satisfied if the infinitessimal work is defined as  
$d'W=mu\circ d(u+v) + (\partial_t+u\partial_x) Vdt$ where the last term, the convective
derivative of the potential energy, agrees with the
definition used in the overdamped limit in \cite{SpeckMehlSeifert08}.

Following~\cite{Celani12} it is convenient to introduce the normalized $n$-dimensional Maxwell-Boltzmann
distribution $W(v,T)= (2\pi T)^{-\frac{n}{2}} \exp\left(-\frac{mv^2}{2T}\right)$
and use $-\frac{mv}{T}\circ dv=-d(\frac{mv^2}{2T})-mv^2\frac{(\partial_t T + (u+v)\cdot\partial_x T)dt}{2T^2}$
to write the functional (\ref{eq:S-ent}) as
\begin{eqnarray}
S_{env} &=& \Delta\log W - \int \partial\cdot u \, dt + \int \frac{v\cdot(f_{eff}-\partial_x T)}{T}dt  \nonumber \\
&-& \int \frac{(mv^jv^i-\mathbf{1}^{ij}T)\partial_j u^i}{T}dt \nonumber \\
&-&\int \left(mv^2-nT \right) \frac{\partial_t T + u\cdot\partial_xT}{2T^2}dt  \nonumber \\
&-&\int \left(mv^2-(n+2)T\right)\frac{v\cdot\partial_xT}{2T^2}dt
\label{eq:S-ent-2}
\end{eqnarray}
This form reflects the eigenvector structure of $M^{\dag}$ since
$M^{\dag}v^iW=-\frac{\gamma}{m} v^iW$, $M^{\dag}(mv^iv^j-\mathbf{1}^{ij}T)W=-2\frac{\gamma}{m} (mv^iv^j-\mathbf{1}^{ij}T)W$
and $M^{\dag}(mv^2-(n+2)T)v^iW=-3\frac{\gamma}{m} (mv^2-(n+2)T)v^iW$.
The expected entropy production $q$ from some time and position in the past up to $(x,v,t)$
in the present fulfills the forward Kolmogorov equation
\begin{equation}
\left(\partial_{t}-M^{\dag}-\epsilon L^{\dag} + \ldots\right) P=C
\end{equation}
where $C$ is the running cost (values at $(x,v,t)$ of the integrands in (\ref{eq:S-ent-2})),
and the multiscale can then be carried out in an analogous manner as above.
To lowest order one finds $P_0=Wq_0(x,t_u,t_f)$ and to
order $\epsilon$ the solvability condition is
$\partial_{t_u}q_0 + \partial_{x_i}(u^iq_0)=-\partial\cdot u$
reflecting a dissipation-less entropy change by the advecting mean flow.
The solution at order $\epsilon$ comprises the same kind of terms as above for the density
and all terms in (\ref{eq:S-ent-2}) (except the first two) counted with the proper eigenvalues
of $(M^{\dag})^{-1}$.
On order $\epsilon^2$ one therefore gets as solvability condition
the terms of a conservation law for a first order spatial entropy function $q_1=\int dv P_1$, \textit{i.e.}
$\partial_{t_u}q_1 + \partial_{x_i}(u^iq_1)$, diffusive
terms for $q_0$, and source terms from the first through fourth lines of (\ref{eq:S-ent-2}).
Up to terms of order $\epsilon^3$ the combinination
$\tilde q_0=q_0+\epsilon q_1$ can hence be seen to obey the forward Kolmorov equation
of the expected value under the process (\ref{eq:overdamped-space-meanflow})
of a combined quantity $\delta S_{env}^{reg}+\delta S_{env}^{quad}+\delta S_{\hbox{anom}}+\delta S_{\hbox{anom}}^{(u)}$
where the first term (contribution from the first line of (\ref{eq:S-ent-2})) is
\begin{equation}
S_{env}^{\hbox{\small reg}} = \int \frac{f^i_{eff}}{T}\left(\circ dX^i-u^idt\right) - \int \partial\cdot u\, dt
\label{eq:S-reg-mean-flow}
\end{equation}
This ``regular entropy production'' is the canonical form of the entropy production of the first-order
stochastic process (\ref{eq:overdamped-space-meanflow}) where the
mean flow ($u$) and the overdamped force $\frac{1}{\gamma}f^i_{eff}$
transform differently under time reversal~\cite{ChG07}.
If the flow is incompressible it also
agrees with the functional derived in~\cite{SpeckMehlSeifert08},
since in the overdamped limit the Maxey terms in $f_{eff}$ are small compared to $f$.
The other three terms in (\ref{eq:S-ent-2})) are hence all ``anomalous''.
The contribution from the second line 
is $\delta S_{\hbox{\small anom}}^{(u)}$ as given in  (\ref{eq:S-anom-u}),
while the similar contribution from the third line is
$\delta S_{env}^{quad}=\int \frac{mn}{4\gamma T^2} (\partial_t T + u\cdot\partial_xT)^2 dt$,
and the contribution from the fourth line 
is $\delta S_{\hbox{\small anom}}$ as given in (\ref{eq:S-anom}).
If the temperature field obeys the advection equation $\partial_t T +  u\cdot\partial_xT = \kappa\partial^2_{xx} T$
and $T$ only varies in space on scale $L$ then $\delta S_{env}^{quad}$
is of higher order and can be ignored.
All the calculations presented above are straightforward 
though somewhat lengthy and
therefore given in Supplementary Material.

{\it The anomalous entropy production of a rotating Brownian particle:}
As an non-trivial illustration that macroscopic thermodynamics correctly predicts the form
of mesoscopic entropy production we will now address the technically more involved case
of a rotating Brownian particle with general moment of inertia and angular velocity friction operators.
The friction coefficient of an object of radius $R$ is by Stokes' law of the order of $R\rho_f \nu$
where $\rho_f$ is the density of the surrounding fluid and $\nu$ its kinematic viscocity. If the
density of the body and the fluid are not too different the relaxation time $t_r$ is hence of order
$R^2/\nu$. This gives another interpretation of the scale separation $t_r/t_f$ (built on $R$) as $D/\nu$.
The Stokes number is $\hbox{St}=\hbox{Re}(R/L)^2$, where $\hbox{Re}=\frac{L\overline{u}}{\nu}$ is the
Reynolds number built on $L$ and $\overline{u}$, and the P\'eclet number is $\hbox{Pe}=\hbox{Re}^{-1}(L/R)^2(\overline{u}/v_{th})^2$.
The advection-diffusion limit is therefore also given by the condition $\overline{u}\sim v_{th}$ and
the limit $\hbox{Re}(R/L)^2\to 0$. We have here
limited ourselves to situations where (at least formally) $R\sim L$ which then means (as an asymptotic limit)
$\hbox{Re}\to 0$.
At very low Reynolds numbers viscous effects are strong and the mean flow is irrotational ($\Omega=\frac{1}{2}\nabla\times u = 0$),
and we can hence treat the rotating Brownian particle as if it were not influenced by the mean flow at all, except
as through how $T$ and $\Gamma$ change in space and time.
For completeness we give in Supplementary Information the full derivation including all terms which would formally appear
if the advection-diffusion limit with no restriction on $\hbox{Re}$; the terms that appear
here can then be found by setting $\Omega$ to zero.

Relaxation on time scale $t_r$ is now described by the operator
\begin{equation}
M = -\frac{\gamma}{m}v^i\partial_{v^i} + \frac{T\gamma}{m^2} \partial_{v^iv^i} - D_{\alpha}^{\beta}\omega^{\alpha}\partial_{\omega^{\beta}} +
TS^{\alpha\beta}\partial_{\omega^{\alpha} \omega^{\beta}}
\label{eq:M-operator}
\end{equation}
where we have introduced the angular velocity friction matrix
$D=I^{-1}\Gamma$ and the angular velocity diffusion matrix
$S=I^{-1}\Gamma I^{-1}$. These two do not necessarily commute, and it is therefore not
always possible to find an orthonormal transformation ($\hat\omega=N\omega$, $N^{-1}=N^t$) which
simultaneously diagonalizes $D$ and $S$.
On the other hand, if the eigenvalues of $D$ are non-degenerate $D$ can always be diagonalized
by a general linear transformation. 
The two operators $D$ and $S$ act on different spaces
and under a general linear transformation they transform as
$\hat D=NDN^{-1}$ and $\hat S=NSN^{t}$.
Therefore, under the further weak assumption that $D$ has full rank
the linear transformation $N$ that diagonalizes $D$ in fact also diagonalizes $S$~\cite{NN-comment}.

Using local charts of the orientations
such that
$Q^i_{j}= \sum_l Q^{(0),i}_{l}\left(\mathbf{1}^l_{j}+\epsilon^l_{jn}\alpha^n +{\cal O}(\alpha^2)\right)$~\cite{Arnold-book}
one can show, in analogy to (\ref{eq:overdamped-space-meanflow}) that the
 overdamped equation of motion of the orientations is
\begin{equation}
d\vec{\alpha}_t = \Gamma^{-1} Q^{-1} M dt + \sqrt{2T\Gamma^{-1}}\circ d\vec{\xi}
\label{eq:overdamped-orientation-no-meanflow}
\end{equation}
where $\sqrt{2T\Gamma^{-1}}$ is the matrix square root, $Q^{-1} M$
the external torque in this frame, and there are no It\^o or spurious corrections
since $\Gamma$ is constant in the body and $T$ does not depend on the orientation.
A globally valid description
such as \textit{e.g.} in terms of Euler angles~\cite{woe62br} (see Supplementary Information)
will contain additional terms depending on the parametrization.
The inertial term in Euler's
equation ($(I^{-1}\omega)\times I\omega$) does not contribute to the overdamped
equation of motion.

The entropy production in the environment of a particle following
(\ref{eq:Langevin Kramers}) and (\ref{eq:wi-simple}) is now the functional
\begin{eqnarray}
S^{\hbox{\small rot}}_{env} &=& \hbox{Eq~(\protect\ref{eq:S-ent})} +\frac{\omega Q^{-1}M}{T}dt -\frac{\omega I}{T}\circ d\omega
\label{eq:S-ent-3}
\end{eqnarray}
which be rewritten as
\begin{eqnarray}
S^{\hbox{\small rot}}_{env} &=& \hbox{Eq~(\protect\ref{eq:S-ent-2})} + \int \frac{\omega Q^{-1}M}{T}dt \nonumber \\
&-&\int \left(\omega I\omega-nT \right) \frac{\partial_t T + u\cdot\partial_xT}{2T^2}dt  \nonumber \\
&-&\int \left(\omega I\omega -nT\right)\frac{v\cdot\partial_xT}{2T^2}dt
\label{eq:S-ent-4}
\end{eqnarray}
The only new terms we need to compute using the relaxation operator (\ref{eq:M-operator})
are $M^{\dag}(\hat\omega_l^2 \hat{I}_{ll} -T)W= -2\hat{D}_{ll}(\hat\omega_l^2 \hat{I}_{ll} -T)W$
and $M^{\dag}(\hat\omega_l^2 \hat{I}_{ll} -T)v^i W= -\left(\frac{\gamma}{m}+2\hat{D}_{ll}\right)(\hat\omega_l^2 \hat{I}_{ll} -T)v^iW$
(no summation over $l$)
and they give terms with the same dependence on temperature gradients and mean flow
as $\delta S_{env}^{\hbox{\small quad}}$ and $\delta S_{env}^{\hbox{\small anom}}$ above.
The second term can be written
$\frac{1}{2T} |\nabla T|^2 \mathbf{Tr}\frac{1}{\gamma\mathbf{1}+2m\mathbf{D}}$ and if the
particle is spherical then $\mathbf{D}=\frac{\gamma_2}{m}\mathbf{1}$ where
$\gamma_2$ has the dimension of a friction coefficient, which gives the simpler expression
quoted in the Introduction.

{\it Discussion:}
In macroscopic Thermodynamics the world is divided up in three parts:
The System, The External System, and the Thermal Environment. For a fluid in a container
the System is the fluid itself and all objects therein, the External System are outside controls
which can remember the previous state of the System, while the Thermal Environment is everything
else which does not remember the previous state of the System.
In Stochastic Thermodynamics~\cite{Sekimoto-book} the System is instead
a Brownian particle or large molecule,
the External System are external controls acting directly on this object, and the Thermal Environment
is the surrounding fluid. This tripartite division of the world
is isomorphic to macroscopic thermodynamics if the time scales of the fluid are much faster than
those of the object, which is reasonable if the object is mesoscopic.
The fluid can thus be close to equilibrium (in the first sense) while
the object is far from equilibrium (in the second sense). The (far from equilibrium) entropy production in
the environment in Stochastic Thermodynamics should then correspond to the (near-equilibrium) change of
entropy per unit time in fluid, which is what we have found in the example
of a Brownian particle with translation and rotation. We believe that this conceptual
distinction might clarify objections to stochastic thermodynamics
and fluctuation relations which have appeared in the literature several times~\cite{CohenMauzerall04,ViRu2008}.


\section*{Acknowledgements}
We thank Ralf Eichhorn for many discussions, Felix Ritort for suggesting
the problem of a mean flow and E.G.D. Cohen for challenging remarks.
E.A. thanks Paolo Muratore-Ginanneschi and Bernard Mehlig for useful comments.
This research is supported by National Natural Science Foundation of
China (Grant No. 11375093), the Ph.D. Program Foundation of
Ministry of Education of China (Grant No.20090002120054),
the Swedish Science Council through grant 621-2012-2982,
and by the Academy of Finland through its Center of Excellence COIN.

\bibliography{fluctuations,LanAurellNotes}%

\begin{thebibliography}{10}
\bibitem{spinning-torque}%
  \BibitemOpen
  \bibinfo {note} {The friction torque acting on a rotating sphere is
  conventionally estimated as $8\pi R^2\eta$}%
  \bibAnnoteFile{NoStop}{spinning-torque}%
\bibitem{Re-crit}%
  \BibitemOpen
  \bibinfo {note} {A Reynolds number for motion in water built on $\overline{u}
  = 1 mm/s$ and $L=1 mm$ is about 1. The critical Reynolds number for a
  Taylor-Couette flow (flow between rotating cylinders) is about 40 if the
  ratio of the cylinder radii is close to one.}%
  \bibAnnoteFile{Stop}{Re-crit}%
\bibitem{order}%
  \BibitemOpen
  \bibinfo {note} {We note $t_u=t_r\epsilon^{-1}$, $L=\ell\epsilon^{-1}$, and
  $u$ has been assumed to be on the order of $v_{th}=\ell/t_r$ hence both sides
  of the conservation law are manifestly of the same order.}%
  \bibAnnoteFile{Stop}{order}%
\end{thebibliography}
\end{document}